# Magnetic phase separation in $La_{1-x}Sr_xCoO_3$ by $^{59}Co$ NMR

P. L. Kuhns[1], M. J. R. Hoch[1], W. G. Moulton[1], A. P. Reyes[1] J. Wu[2] and C. Leighton[2]
[1]National High Magnetic Field Laboratory [2]Department of Chemical Engineering and Materials Science, University of Minnesota

$^{59}Co$ NMR measurements on $La_{1-x}Sr_xCoO_3$ reported here establish unequivocally, for the first time, the coexistence of ferromagnetic regions, spin glass regions, and hole poor low spin regions at all x values from 0.1 to 0.5. A zero external field NMR spectrum, which is assigned to the ferromagnetic regions, has a spectral shape which is nearly x independent at 1.9 K, as are the relaxation times, $T_1$ and $T_2$. The integrated spectral area increases rapidly with x up to x=0.2 and then becomes almost constant for larger x. In a field of 9.97 T, a narrow NMR line is observed at 102 MHz, identical to that found in x=0 samples in previous work. The integrated intensity of this spectrum decreases rapidly with increasing x, and is ascribed to hole poor low spin regions. Beneath this spectrum, a third broad line, centered at 100 MHz, with shorter $T_1$ and $T_2$, is assigned to a spin glass, intermediate spin phase.

There is growing evidence that inhomogeneities are crucial for understanding the unusual magneto-electronic properties of doped perovskites including manganites and cuprates[1]. It is likely that magnetic phase separation is at least as important in the cobaltites as in other oxides, as evidenced by the unusual magnetic properties of doped cobaltites which are interpreted in terms of magnetic clusters [2]. The present work is directed toward identifying and investigating such phase separation in $La_{1-x}Sr_xCoO_3$ using NMR as a microscopic probe. The magnetic and transport properties of the doped perovskite systems $La_{1-x}Sr_xMO_3$ (M= Mn or Co) have attracted considerable recent attention [1,2,3,4]. Hole doping by substitution of a divalent ion for La induces a metal-insulator (MI) transition together with ferromagnetism (FM), while double exchange, involving Mn or Co ions in different charge states, provides the FM coupling mechanism.

Magnetic and electrical transport measurements on $La_{1-x}Sr_xCoO_3$ (LSCO) have revealed spin-glass (SG) properties, large magnetoresistance and giant anomalous Hall effects which are not well understood.[5,6,7]. A recently proposed phase diagram [7], Fig. 1, shows paramagnetic, SG and ferromagnetic cluster glass (CG) regions with the low temperature SG to FM CG change occurring at x=0.18, coincident with a MI transition. For Sr concentrations in the range 0.2< x<0.5, the transition temperature, $T_c$, to a FM phase is weakly x dependent and approximately 240 K The CG magnetic properties include memory effects characteristic of glassy systems and are in some ways similar to those of reentrant spin-glass (RSG) systems [8,9,10,11,12]. LSCO has a rhombohedrally distorted perovskite structure which is close to cubic [2]. Recent experimental [7] and theoretical work [13] suggests that both $Co^{3+}$ and $Co^{4+}$ ions are in the intermediate spin-state (IS) $t^5_{2g} e^1_g$ for $Co^{3+}$ and $t^4_{2g} e^1_g$ for $Co^{4+}$. There is evidence of a small lattice distortion [2] of the oxygen octahedron, involving changes in the Co-O bond angle near the ordering temperature for x≥0.1. Small changes in covalent hybridization of Co d and O p states can produce large changes in properties. In this mixed valence system, double exchange between the $Co^{3+}$ and $Co^{4+}$ give rise to FM coupling, while exchange between like ions ($Co^{3+}$ or $Co^{4+}$) produce antiferromagnetic (AFM) coupling. AFM and FM competition between single charge state regions and mixed valence regions leads to frustration and SG properties. The MI transition occurs when the percolation threshold of FM clusters is reached, and is coincident with the onset of FM-like behavior in the bulk magnetic properties.

There are extensive experimental and theoretical efforts underway to understand the underlying mechanisms of magnetic phase separation [14]. In the manganites neutron diffraction [15], scanning tunneling microscopy [16], noise measurements [17] and NMR [18,19] have been used to probe magnetic phase separation, while x-ray absorption fine structure [20] has revealed a possible structural precursor. For LSCO, high resolution transmission electron microscopy has revealed an inhomogeneous distribution of Sr-rich and La-rich regions [2] with sizes 8 to 40 nm. Neutron diffraction experiments [2] show that the low temperature FM correlation length is several hundred angstroms, while small-angle-neutron-scattering (SANS) shows magnetic clustering effects, on a scale of 10 to 15 $A^0$.

There have been only a few previous NMR publications on the cobaltites. Bose et al's [21] low field continuous wave (8MHz) $^{59}Co$ (I=7/2) NMR measurements for the $LaCoO_3$ system gave a quadrupole coupling constant, $\nu_Q$ = 0.6 MHz with axial symmetry (η =0). Itoh and Natori's [22] later $^{59}Co$ spin echo experiments on $La_{1-x} Sr_x Co O_3$ ( x ≤0.15 ) confirmed this $\nu_Q$ value. For Sr doped samples, the line broadened, and became unobservable at x=0.15. At 4.2 K, a broad zero field spectrum, observed for x≥0.2, was interpreted in terms of FM regions. Kobayashi et al. [23] have very recently made $^{59}$Co spin echo measurements at 95.0 MHz on single crystals of $LaCoO_3$ and $La_{0.95}Sr_{0.05}CoO_3$, with results which are consistent with the earlier work.

In our work, both zero field and high field $^{59}$Co NMR have been employed to obtain information on the microstructure and magnetic phase separation of LSCO over a wide x range. The samples were prepared by solid–state reaction at 980 $^o$C followed by cold pressing, sintering for 24 hours at 1350 $^o$C and slow cooling. Characterization using X-ray diffraction, scanning electron microscopy, iodometric titration and high resolution energy dispersive X-ray spectroscopy showed that the material is stochiometric and uniform, within the resolution, with grain sizes of microns [7].

Pulsed NMR measurements were made using a computer controlled wide-band spectrometer, designed and built at the NHMFL, with an untuned low temperature probe. Spin echo signals were observed either in zero field or in fields up to 10 T . The spectrometer response and calibration were carried out over the range 40- 250 MHz. using a $^7$Li NMR signal. The blank probe at 1.9 K gave no signal over the frequency range used. No measurable differences in the spectra were found for FC or ZFC protocols. The signal was monitored during cooldown (300-1.9 K), with cooldown times ranging from 16 hours to less than 10 minutes. The time to reach maximum signal amplitude after reaching base temperature, 1.9 K, was of order hours for slow cooling and 30 minutes for fast cooling. For some cooldowns, a sudden increase in signal amplitude was observed with the signal roughly doubling in size in less than 2-3 minutes.. Although this suggests some cooperative expansion of the FM regions, it is important to note the signal intensity and integrated area came to the same *final value* independent of the cool down rate or behavior. All spectra have a broad peak centered at 173 MHz, with unresolved structural features at other frequencies, as shown in Fig. 1(a). The data have been corrected for the measured spectrometer response and scaled with the molar fraction in each case. For NMR in FM, the RF excitation pulse as well as the signal are enhanced by the hyperfine field within domains, and additionally enhanced within any domain walls. The enhancement factor is defined by $\eta=B_{eff}/B_{RF}$, where $B_{eff}$ is the RF field seen by the nucleus, and $B_{RF}$ is the applied radio frequency field. Typical values of $\eta$ in the manganites [24] are 100-3,000 for nuclei in domains, and 5,000 to 30,000 within domain walls. $\eta$ for $^{59}$Co NMR was about 40, independent of composition, a low value compared to that found in the FM manganites. The spectral shape is relatively independent of x, but with a small increase in the spectral component centered at 120 MHz, as x is increased. The pulse lengths used in the spin echo measurements were 0.6 μs, covering a spectral width greater than 1MHz. The ZF detection of FM NMR for x<0.18 implies that the region denoted SG or CG in the proposed phase diagram [7] contains FM regions below the percolation limit. The Hamiltonian for the system in zero field is:

$$H = H_{hf} + H_Q$$

where $H_{hf} = \mathbf{I.A.S}$ is the hyperfine coupling, $H_Q$ is the quadrupolar Hamiltonian, $\mathbf{I}$ is the nuclear spin, and $\mathbf{A}$ is the hyperfine coupling tensor including both the isotropic (contact) part and the anisotropic part,. The quadrupolar interaction is a small perturbation on $H_{hf}$ . For spins I with hyperfine coupling $A_i$ the resonance frequency is $\nu=\gamma_n A_i <S>$ , where $\gamma_n$ is the nuclear magnetogyric ratio and $<S>$ the ensemble average electron spin along the principal axis. We assume that the magnetization, $M=Ng\beta<S>$, where N is the number of spins, is uniform within the FM regions and the lineshape is attributed to a distribution of hyperfine couplings. In the metallic phase (x>0.18) and below the percolation threshold (0.10<x<0.18), the FM cluster regions are itinerant FM. Double exchange processes on a timescale h/J, where J is the exchange coupling, lead to an averaging of the hyperfine coupling between $Co^{3+}$ and $Co^{4+}$ sites. This gives rise to an averaged spin state for these ions. It is likely that local lattice distortions, due to the random replacement of La ions by larger Sr ions, lead to changes in the d-orbitals for adjacent Co ions, producing changes in the core polarization and hence the local hyperfine coupling, giving a broad range of hyperfine couplings. The lineshape fit with multiple Gaussian peaks is shown in Fig 2(b). Four Gaussians give an excellent fit. Three Gaussians give a poor fit, while non physical parameters occur with five. Attempts to fit the spectrum using the same fixed width, for as many as six Gaussians, were unsuccessful. This indicates that the different regions within the sample have different distributions of hyperfine couplings. Obviously, this fit is not unique, but provides some indication of the underlying hyperfine distribution. The area of the peak centered at 172 MHz is essentially independent of x, while the area of the low frequency peak increases rapidly with x, reflecting the dependence of the Co hyperfine couplings on the number of nearest neighbor Sr ions. The high frequency Gaussian component correspondingly decreases with increasing x.

To investigate the multiphase character of LSCO, spin echo NMR experiments were carried out for samples with a range of x values in an applied magnetic field of 9.97 T, for which the unshifted $^{59}$Co Larmor frequency is 100 MHz. Spectra are shown in Fig. 3. All the spectra have two main components, a narrow line centered at 102 MHz, with the same Knight shift as in LaCoO$_3$, and a very broad component centered at 100 MHz. The narrow 102MHz spectrum is assigned to hole poor regions given the similarity to data on x=0 samples. As x is increased, the 100 MHz line develops a low frequency tail which extends over a range of more than 10 MHz, with an enhancement factor of 2. A two Gaussian fit is shown in the Fig. 3 inset. For all x, $T_1$ is 820 msec and $T_2$ is 247 µsec at 2 K, similar to the values for LaCoO$_3$ [23]. The $T_1$ recovery curves require a stretched exponential fit indicating a distribution of relaxation times. This is at least partially due to admixture of both the narrow (hole poor) and broad (glass) portions of the spectrum. $T_2$ is single exponential.

$^{59}$Co Knight shift measurements in undoped and lightly doped LaCoO$_3$ have revealed a low temperature value of <K>=0.02, which is attributed to van Vleck orbital paramagnetism for Co ions in the LS (S=0) state [22]. As the temperature is raised to around 100 K, <K> increases. This is attributed to a transition of the ions to the IS ( S=1), state with an energy gap $\Delta \cong 180$ K, due to the small difference between the intra-atomic exchange energy and the crystal field energy. The 102 MHz line observed in the present high field work at 1.9 K is consistent with the Knight shift of 0.02 observed in x=0 samples. It is therefore concluded that this spectrum arises from ions in the LS Co$^{III}$ state in hole poor regions, possibly of very low Sr concentration. Our results suggest that a number of LS ions persist in samples with x values as high as 0.4. It has been reported that annealing at $\geq 1200$ °C suppresses this phase [25]. However, these samples were annealed at 1350 °C, and the low spin phase is clearly present at all x values. This suggests the possibility that the presence of a low spin phase may be intrinsic to the system.

Fig. 4 shows the area under the zero field and 9.97T NMR spectra, proportional to the number of Co spins in FM and in non-FM hole- poor regions, respectively, as a function of x. There is a marked increase in FM area as x increases from 0.1, followed by a plateau region for $x>x_c$. The smooth increase in (FM) spectral area through the MI transition is noteworthy.

Measurements of the spin-lattice, $T_1$, and spin-spin, $T_2$, relaxation times have been made as a function of temperature for the FM component in zero field and will be reported on in detail elsewhere. The rates $1/T_1$ and $1/T_2$ depend linearly on T and are, within experimental error, independent of Sr concentration. The ratio $T_2 / T_1$ is approximately constant over the temperature range covered, suggesting a common mechanism for the two processes. $T_2$ becomes so short as T is raised that measurements at temperatures higher than 25 K are not possible. There was no change in $T_1$ or $T_2$ in crossing the MI percolation limit. These results show that the FM behavior is not closely coupled to the MI percolation transition in this system. The near independence of the hyperfine coupling and the relaxation times on x implies that the FM clusters have essentially the same ordering properties and spin dynamics above and below the MI (percolation) transition. The area under the 102 MHz line (the low spin component) decreases rapidly with x, as shown in Fig. 4, but with the signal observable for x values as high as 0.4. Similarly, the 100 MHz peak decreases in magnitude with x but the relatively poor signal-to-noise ratio for $x>x_c$, together with the evolution of the long tail region, makes determination of the area impractical.. Because of their proximity to the unshifted $^{59}\gamma$ value and η near 1, these signals must come from the low spin regions of the samples. The data of Fig. 4 provides clear and unequivocal evidence for the multiphase character of the material— *FM and non-FM regions coexist over the entire composition range studied*. As a result of uncertainties in signal enhancement effects and other factors, it is not a simple matter to reliably compare the number of spins in the FM and non-FM regions. It is, however, clear from the results that, for x>0.18, the ratio of FM to non-FM spins becomes large.

When a magnetic field in the range 0.5 to 5.0 T was applied to the system, the peak spectral amplitude observed with short pulses (0.6 µs), corresponding to the signal with the largest η, showed a rapid decrease as the field was increased, although the spectral lineshape remained essentially unaltered. For B= 5.0 T, the signal amplitude was unobservably weak at T=2 K. The lack of shift with field is reminiscent of the shielding within domain walls in a conventional FM. Using longer pulses (5 µs), weak echo signals were observed with a shift of the spectrum to higher frequencies with increasing field corresponding to the hyperfine field acting parallel to the magnetization direction, in contrast with the usual shift to lower frequencies found in other FM cobalt systems. Further work is needed to understand the behavior of the FM component of the system in applied fields

These $^{59}$Co NMR measurements clearly establish the inhomogeneous magnetic nature of LSCO, with FM, spin glass, and LS regions coexisting on both the insulating and metallic sides of the MI transition. Moreover, the data demonstrate that the FM clusters have the same couplings and spin dynamics, within the time scale of the NMR measurements, above and below the MI (percolation) transition. The large zero field linewidth and single central peak support a single average spin state for the Co ions in FM regions with a large distribution of hyperfine couplings. Taken together with the 102 MHz high field NMR signal amplitude, which decreases with x, the results are consistent with intermediate spin (FM plus SG) and low spin states occurring in different spatial regions of the samples. The question of whether the S=0 part of the sample arises from Sr poor regions, due to intrinsic chemical or structural phase separation, or hole poor regions from some other mechanism is currently unclear.

Partial support by the National Science Foundation under cooperative agreement DMR-0084173 and the State of Florida is gratefully acknowledged..

References
1. E. Dagotto, T Hotta, and A. Moreo Phys. Rep.**344**, 1 (2001).
2. R. Caciuffo et al., Phys. Rev. **B 59**, 1068 (1999).
3. J.M.D. Coey, M. Viret and S. von Molnár, Adv. in Phys. **48** 167 (1999).
4. S.Jin, T.H. Teifel, Science **264**, 413 (1994) ; Y, Tokura, N. Nagaosa,. Science **288**, 462 (2000).
5. G. Briceno , H. Chang ,X. Sun, P.G. Schultz ,X.-D. Xiang, Science **270**, 5234 (1995).
6. M. Itoh et al., J. Phys. Soc. Japan, **63**, 1486 (1994).
7. J. Wu and C. Leighton ( submitted to Phys.Rev. B).
8. M. Itoh, I. Natori, S. Kubota , K. Motoya, J. Phys. Soc. Japan, **63**, 1486 (1994).
9. G. Aeppli, S. M. Shapiro, R. J. Bergeneau, and H. S. Chan, Phys. Rev. **B 28**, 5160 (1983).
10. H. Maletta nd P. Convert, Phys. Rev. Lett., **42**, 108 (1979) .
11. D.N.H. Nam et al , Phys.Rev. **B 59**, 4189 (1999).
12. Yamaguchi et al., J. Phys. Soc. Japan, **64**, 1885 (1995).
13. P. Ravindran et al. , J. Appl. Phys., **91**, 291 (2002).
14. J. Burgy et al., Phys. Rev. Lett., **87**, 277202-1 (2001).
15. J. W. Lynn, et al., Phys. Rev. Lett. **76**, 4046 (1996).
16. M. Fath et al., Science **285**, 1540 (1999).
17. B. Raquet, Phys. Rev. Lett **84**, 4485 (2000).
18. G.Papavissiliou et al., Phys. Rev. Lett. **84**, 761 (2000).
19. M.M. Savosta et al., Phys. Rev. Lett. **87**, 137204 (2001).
20. T. Shibata et al., Phys. Rev. Lett. **84** 207205 (2002).
21. M. Bose et al., Phys. Rev. **B 26**, 4871 (1982).
22. M. Itoh, I. Natori, J. Phys. Soc. Japan **64**, 970 (1995).
23. Y.Kobayashi et al.,Phys. Rev. **B 62**, 410 (2000).
24. M.M. Savosta, V.A.Borodin, P. Novak, Phys. Rev. **B 59**, 8778 (1999).
25. P. S. Anil Kumar, P. A. Joy, S. K. Date, J. Appl. Phys. **83**, 7375 (1998).

Figure captions

Figure 1    Phase diagram of $La_{1-x}Sr_xCoO_3$ according to reference 7.

Figure 2    (a)  Zero field $^{59}Co$ NMR spectrum for $La_{1-x}Sr_xCoO_3$ for x=0.1, 0.18, 0.2, and 0.4. (b)Four Gaussian fit of the 0.3 spectrum. The peaks are at 206, 198, 172, and 142 MHz. The peak positions are independent of x, but the weight of the low frequency peak increases at the expense of the two high frequency peaks as x increases.

Figure 3   $^{59}Co$ NMR specra in field B=9.97 T for Sr concentrations x in the range 0.1. to 0.4. The fitted Gaussian curves are centered at 100 Mhz and 102 MHz , respectively.

Figure 4    The integrated area of the zero field spectra (right scale) and the LS 9.97 T NMR spectra (left scale) as a function of x. The sum of areas of the Gaussian fits gave identical values. For the zero field spectra, the weight of the low frequency Gaussian increases with x at the expense of the high frequency Gaussian.

PACS: 75.30.Kz
       75.50.Cc
       76.60.Jx

.

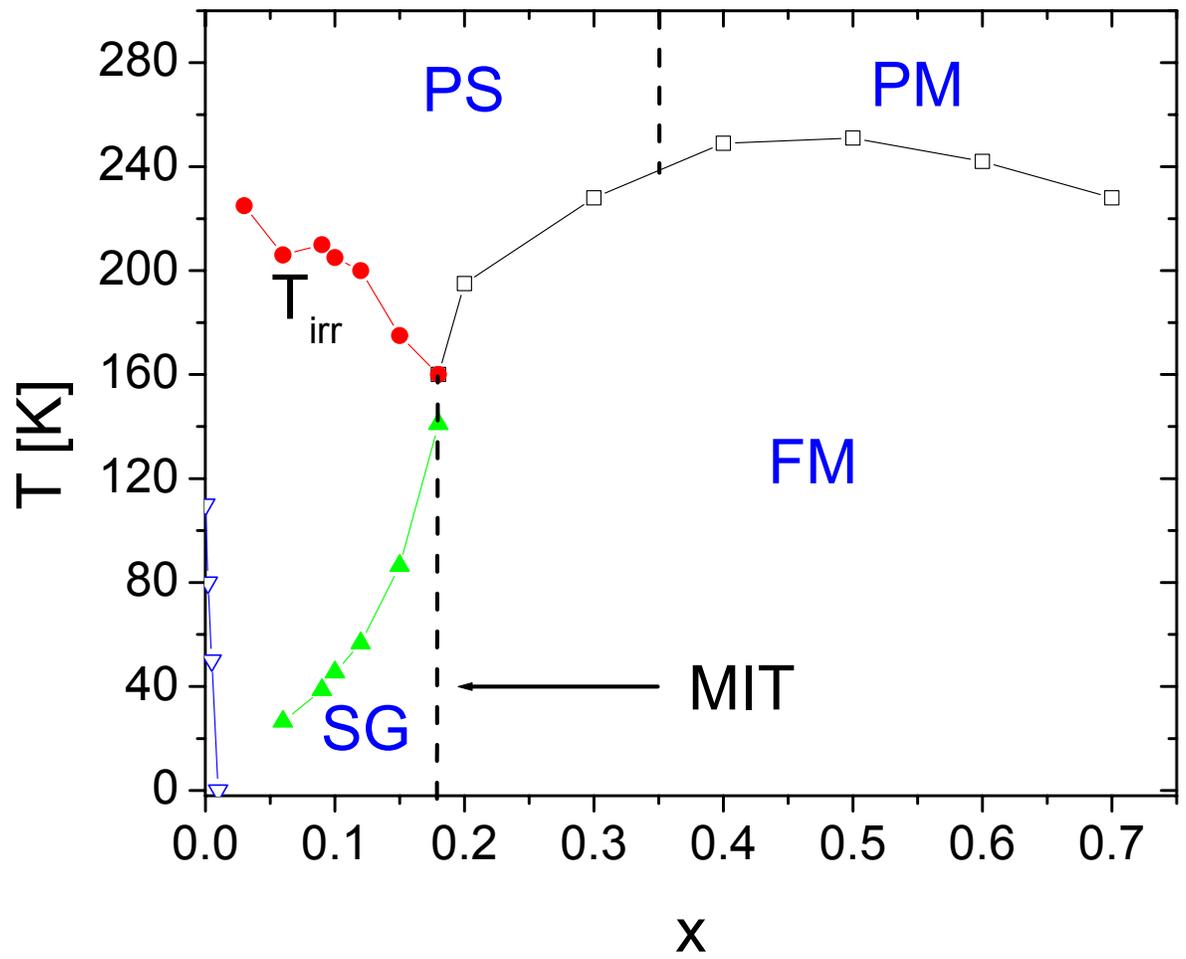

Figure 1 Kuhns Phys. Rev. Lett

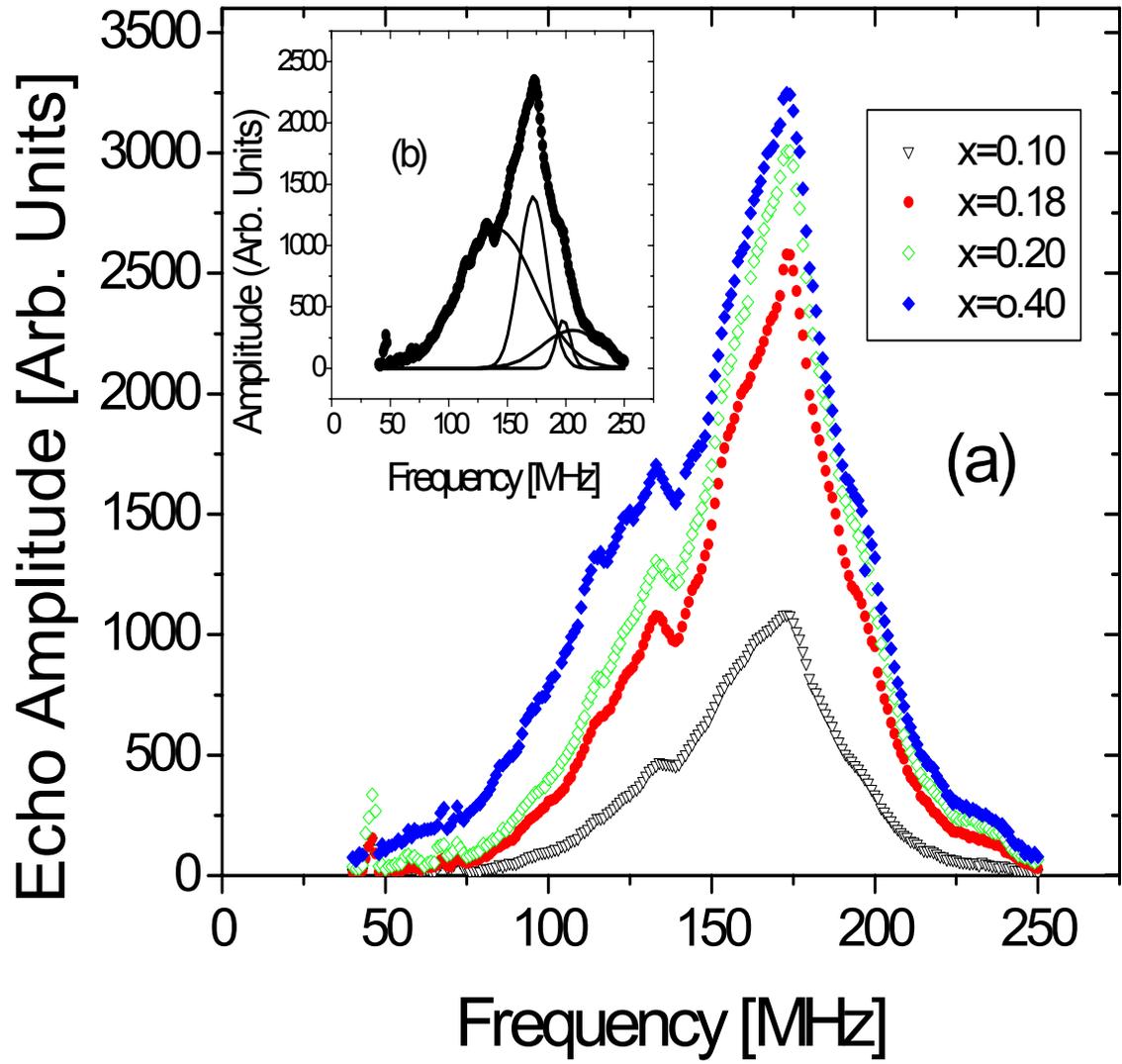

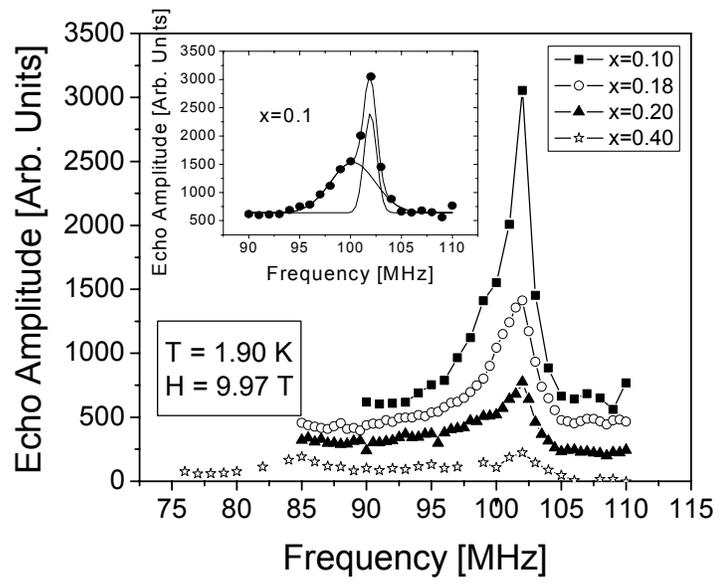

Fig.3

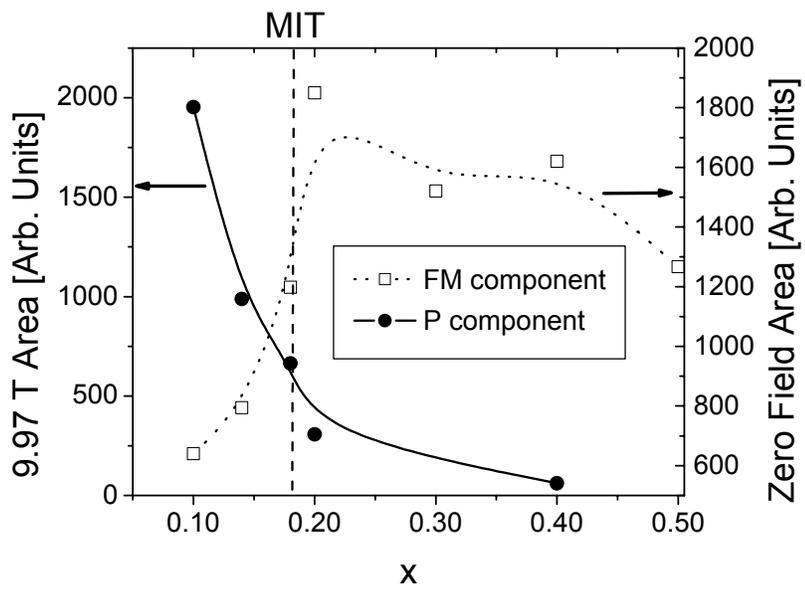

Fig. 4